\documentstyle{article}
\newcommand{\f}{\begin{equation}}
\newcommand{\ff}{\end{equation}}
\newcommand{\tr}{{\rm tr}}
\begin{document}

\title{\bf Von Neumann algebra automorphisms \\ and
time-thermodynamics relation \\
in general covariant quantum theories}
\author{A. Connes${}^1$, \  C. Rovelli${}^2$\\
${}^1$ Institut des Hautes Etudes Scientifiques, \\
35 route de Chartres, 91440 Bures sur Yvette, France. \\
${}^2$ Physics Department, University of Pittsburgh, \\
Pittsburgh,
Pa 15260, USA, \\
and Dipartimento di Fisica Universita' di Trento, Italia.}
\date{\today}
\maketitle

{\bf  Abstract}
\vskip.2cm

We consider the cluster of problems raised by the relation
between the notion of time, gravitational theory, quantum theory
and thermodynamics;  in particular, we address the problem of
relating the "timelessness" of the hypothetical fundamental
general covariant quantum field theory with the "evidence" of the
flow of time.  By using the algebraic formulation of quantum
theory, we propose a unifying perspective on these problems,
based on the hypothesis that in a generally covariant quantum
theory the physical time-flow is not a universal property of the
mechanical theory, but rather it is determined by the
thermodynamical state of the system ("thermal time hypothesis").
We implement this hypothesis by using a key structural property
of von Neumann algebras: the Tomita-Takesaki theorem, which
allows to derive a time-flow, namely a one-parameter group of
automorphisms of the observable algebra, from a generic thermal
physical state.  We study this time-flow, its classical limit, and
we relate it to various characteristic theoretical facts, as the
Unruh temperature and the Hawking radiation. We also point out
the existence of a state-independent notion of "time", given by
the canonical one-parameter subgroup of outer automorphisms
provided by the Cocycle Radon-Nikodym theorem.

\newpage

{\bf 1. Introduction}
\vskip.3cm

The relations between time, gravity, thermodynamics and
quantum theory form a cluster of unsolved problems and puzzling
surprising theoretical facts.  Among these there are the much
debated "issue of time" in quantum gravity
\cite{libri,isham,time}, the
lack of a statistical mechanics of general relativity
\cite{statistical}, and the elusive thermal features of quantum
field theory in curved spaces, which manifest themselves in
phenomena as the Unruh temperature \cite{unruh} or Hawking
black hole radiation \cite{hawking}.  It is a common opinion that
some of these facts may suggest the existence of a profound
connection between general covariance, temperature and quantum
field theory, which is not yet understood.   In this work we
discuss a unifying perspective on this cluster of problems.

Our approach is based on a key structural property of von Neumann
algebras.  The links between some of the problems mentioned and
central aspects of von Neumann algebras theory have already been
noticed.   A prime example is the relation between the KMS theory
and the Tomita-Takesaki theorem \cite{haag}.  Rudolf Haag
describes this connection as ``a beautiful example of
`prestabilized harmony' between physics and mathematics"
(\cite{haag}, pg. 216).  Here, we push this relation between a deep
mathematical theory and one of the most profound and unexplored
areas of fundamental physics much further.

The problem we consider is the following. The physical
description of systems that are not generally covariant is based
on three elementary physical notions: observables, states, and
time flow.  Observables and states determine the kinematics of
the system, and the time flow (or the 1-parameter
subgroups of the Poincare' group) describes its dynamics. In
quantum mechanics as well as in classical mechanics, two
equivalent ways of describing the time flow are available: either
as a flow in the state space (generalised Schr\"odinger picture),
or as a one parameter group of automorphisms of the algebra of
the observables (generalised Heisenberg picture).  In classical
Hamiltonian mechanics, for instance, the states are represented
as points $s$ of a phase space $\Gamma$, and observables as
elements $f$ of the algebra $A=C^\infty(\Gamma)$ of smooth
functions on $\Gamma$.  The hamiltonian, $H$, defines a flow
$\alpha^S_t: \Gamma\rightarrow\Gamma,$ for every real $t$, on
the phase space (generalised Schr\"odinger picture), and,
consequently, a one parameter group of automorphism $(\alpha_t
f)(s)=f(\alpha^S_t s))$ of the observable algebra $A$
(generalised Heisenberg picture).

This picture is radically altered in general covariant theories (as
general relativity [GR from now on], or any relativistic theory
that incorporates the gravitational field, including, possibly, a
background independent string theory).  In a general covariant
theory there is no preferred time flow, and the dynamics of the
theory cannot be formulated in terms of an evolution in a single
external time parameter.   One can still recover weaker notions of
physical time:  in GR, for instance, on any given solution of the
Einstein equations one can distinguish timelike from spacelike
directions and define proper time along timelike world lines.
This notion of time is weaker in the sense that the full dynamics
of the theory cannot be formulated as evolution in such a
time.\footnote{Of course one should avoid the unfortunate and
common confusion between a dynamical theory {\it on\ } a given
curved geometry with the dynamical theory {\it of\ } the
geometry, which is what full GR is about, and what we are
concerned with here.}    In particular, notice that this notion of
time is {\it state dependent}.

Furthermore, this weaker notion of time is lost as soon as one
tries to include either thermodynamics or quantum mechanics
into the physical picture, because, in the presence of thermal or
quantum ``superpositions" of geometries, the spacetime causal
structure is lost.  This embarrassing situation of not knowing
``what is time" in the context of quantum gravity has generated
the debated issue of time of quantum gravity.   As emphasized in
\cite{statistical}, the very same problem appears already at the
level of the classical statistical mechanics of gravity, namely as
soon as we take into account the thermal fluctuations of the
gravitational field.\footnote{The remark of the previous note
applies here as well. Thermodynamics in the context of dynamical
theories {\it on\ } a given curved geometry is well understood
\cite{tolman}.}   Thus, a basic open problem is to understand how
the physical time flow that characterizes the world in which we
live may emerge from the fundamental ``timeless" general
covariant quantum field theory \cite{abhaylh}.

In this paper, we consider a radical solution to this problem. This
is based on the idea that one can extend the notion of time flow
to general covariant theories, but this flow depends on the
thermal state of the system.   More in detail, we will argue that
the notion of time flow extends naturally to general covariant
theories, provided that: i. We interpret the time flow as a 1-
parameter group of automorphisms of the observable algebra
(generalised Heisenberg picture); ii. We ascribe the temporal
properties of the flow to thermodynamical causes, and therefore
we tie the definition of time to thermodynamics; iii. We take
seriously the idea that in a general covariant context the notion
of time is not state-independent, as in non-relativistic physics,
but rather depends on the state in which the system is.

Let us illustrate here the core of this idea -- a full account is
given in sec. 3 below.  Consider classical statistical mechanics.
Let $\rho$ be a thermal state, namely a smooth positive
(normalized) function on the phase space, which defines a
statistical distribution in the sense of Gibbs \cite{Gibbs}.  In a
conventional non-generally covariant theory, a hamiltonian $H$ is
given and the equilibrium thermal states are Gibbs states $\rho =
exp\{ - \beta H\}$.   Notice that the information on the time flow
is coded into the Gibbs states as well as in the hamiltonian.
Thus, the time flow $\alpha_t$ can be recovered from the Gibbs
state $\rho$  (up to a constant factor $\beta$, which we
disregard for the
moment).   This fact suggests that in a thermal context it may
be possible to ascribe the dynamical properties of the system to
the thermal state, rather than to the hamiltonian:  The Gibbs
state determines a flow, and this flow is precisely the time
flow.

In a general covariant theory, in which no preferred dynamics and
no preferred hamiltonian are given, a flow $\alpha_t^\rho$, which
we will call the {\it thermal time of\,} $\rho$, is determined by
{\it any\,} thermal state $\rho$.  In this general case, one can {\it
postulate}  \cite{statistical} that {\it the thermal time
$\alpha_t^\rho$ defines the physical
time.}

We obtain in this way a general {\it state dependent\ } definition
of time flow in a general covariant context.  If the system is not
generally covariant and is in a Gibbs state, then this postulate
reduces to the Hamilton equations, as we shall show.  In the
general case, on the other side, concrete examples show that the
postulate leads to a surprisingly natural definition of time in a
variety of instances \cite{statistical2}.  In particular, the time
flow determined by the cosmological background radiation
thermal state in a (covariantly formulated) cosmological model
turns out to be precisely the conventional
Friedman-Robertson-Walker time \cite{statistical2}.   In other
words, we describe the universe we inhabit by means of a
generally covariant theory without a preferred definition of time,
but the actual thermal state that we detect around us and the
physical flow that we denote as time {\it are\,} linked by the
postulate we have described.

The fact that a state defines a one parameter family of
automorphisms is a fundamental property of von Neumann
algebras.  The relation between a state $\omega$ over an algebra
and
a one parameter family of automorphisms $\alpha_t$ of the
algebra is the content of the Tomita-Takesaki theorem and is
at the roots of von
Neumann algebra classification, and therefore at the core of von
Neumann algebra theory \cite{al}.   The link between a thermal
state and a
time flow described above can be seen as a special case of such a
general relation.  This observation opens the possibility of widely
extending the application of the idea described above, and to
relate this idea to powerful mathematical results on the one side,
and to the thermal properties of accelerated states in quantum
field theory on the other side:

The observables of a quantum system form a $C^\star$-algebra.
States are positive linear functionals over the algebra.
In a non-generally covariant theory, the definition of the theory
is completed by the hamiltonian, or, equivalently, by a
representation of the Poincare' group.   In a generally covariant
theory, on the other side, we have only the algebra of the gauge
invariant observables and the states \cite{time}.  Given a state,
the Tomita-Takesaki theorem provides us with a 1-parameter
group of automorphisms $\alpha_t$ of the weak closure of the
algebra, the modular
group.   Thus, we may extend the thermal time postulate to the
quantum theory, by assuming that
\begin{itemize}
\item the physical time is the modular flow of the thermal state.
\end{itemize}
We obtain a state dependent definition of the
physical time flow in the context of a generally covariant {\it
quantum\,} field theory.   We will show in sec. 4 that the
classical limit of the flow defined by the modular group is
the flow considered above in the classical theory.

One of the consequences of this assumption is that the puzzling
thermal properties of quantum field theory in curved space,
manifested in particular by the Unruh and the Hawking effects,
appear in a completely new light. In fact, we shall show in sec. 4
that they can be directly traced to the postulate.  A second
consequence is more subtle. A key result in von Neumann algebras
theory is the Cocycle Radon-Nikodym theorem \cite{al},
which implies that the
modular flow {\it up to inner automorphisms, } is an intrinsic
property of the algebra, independent from the states.  In this
subtle sense a von Neumann algebra is intrinsically a ``dynamical"
object.   This result is at the core of the von Neumann algebra
classification.  By interpreting the modular flow as the physical
time flow, this result assumes a deep physical significance: it is
the intrinsic algebraic structure of the observable algebra that
determines the allowed ``time flows".   We describe these
consequences of our postulate in sec. 4.

The problem of constructing a consistent generally
covariant quantum field theory, with a reasonable physical
interpretation, represents the key problem of quantum gravity.
The postulate we introduce here is a tentative step
in this direction.   It addresses the issue of connecting a
generally covariant quantum structure with the observed physical
time evolution.   Moreover, it provides a unified perspective on a
variety of open issues that space from the possibility of defining
statistical mechanics of the gravitational field to the Unruh
and Hawking effects.  The aim of this paper aims is solely to
introduce this postulate and to describe the main ideas and
mathematical ingredients on which it is based.   We leave an
extensive analysis of its physical consequences to future work.

In the following sections, we begin by recalling the mathematical
results on which our discussion is based, and the present status
of the problem of the selection of a physical time in GR (sec.2);
we then discuss the main idea in detail (sec.3), and its main
consequences (sec.4).  Sec.5 overviews the general perspective
that we are presenting.
\vfill\pagebreak

\vskip.9cm
{\bf 2. Preliminaries}
\vskip.3cm

We begin by reviewing few essential fact from mathematics and
from the physics of general covariant theories.  In sect.2.1, we
recall some central results on von Neumann algebras. Detailed
introductions to this area of mathematics and proofs can be found
for instance in refs.\cite{haag,al,algebre}.  In sec.2.2, we discuss
the notion of physical time in GR.   For the reader familiar with
the issue of time in gravity, sec.2.2 has the sole purpose of
declaring our background conceptual assumptions.  We have no
presumption that the conceptual framework in terms of which we
define the problem is the sole viable one.

\vskip.3cm
{\it 2.1 Modular automorphisms}
\vskip.3cm

A concrete $C^\star$-algebra is a linear space $\cal A$ of bounded
linear operators on a Hilbert space $\cal H$, closed under
multiplication, adjoint conjugation (which we shall denote as
${}^\star$), and closed in the operator-norm topology.  A concrete
von Neumann algebra is a $C^\star$-algebra closed in the weak
topology.    A positive operator $\omega$ with unit trace on the
Hilbert space $H$ (in quantum mechanics: a density matrix,
or a physical state) defines a normalized positive linear
functional over $\cal A$ via
\f
	\omega(A) = Tr[A \omega]. 			\label{state}
\ff
for every $A\in{\cal A}$.  If $\omega$ is (the projection operator
on) a ``pure state" $\Psi\in H$, namely if
\f
	\omega = |\Psi\rangle\langle\Psi|
\ff
(in Dirac notation) then eq.(\ref{state}) can be written as the
quantum mechanical expectation value relation
\f
	\omega(A) = \langle\Psi|A|\Psi\rangle.
\ff

An abstract $C^\star$-algebra, and an abstract von Neumann
algebra (or $W^\star$-algebra), are given by a set on which
addition, multiplication, adjoint conjugation, and a norm are
defined, satisfying the same algebraic relations as their concrete
counterparts \cite{algebre}.  A state $\omega$ over an abstract
$C^\star$-algebra $\cal A$ is a normalized positive linear
functional over $\cal A$.

Given a state $\omega$ over an abstract $C^\star$-algebra $\cal
A$, the well known Gelfand-Naimark-Segal construction provides
us with a Hilbert space $H$ with a preferred state $|\Psi_0\rangle$,
and a representation $\pi$ of $\cal A$ as a concrete algebra of
operators on $H$, such that
\f
	\omega(A) = \langle \Psi_0 | \pi(A) | \Psi_0 \rangle.
\ff
In the following, we denote $\pi(A)$ simply as $A$.  Given
$\omega$ and the corresponding GNS representation of
$\cal A$ in $H$, the set of all the states $\rho$ over $\cal A$
that can be represented as
\f
			\rho(A) = Tr[A \rho ]
\ff
where $\rho$ is a positive trace-class operator in $H$, is denoted
as the folium determined by $\omega$.    In the
following,  we shall consider an abstract $C^\star$-algebra $\cal
A$, and a preferred state $\omega$.  A von Neumann algebra $\cal
R$ is then determined, as the closure of $\cal A$ under the weak
topology determined by the folium of $\omega$.

We will be concerned with 1-parameter groups of automorphisms
of a von Neumann algebra $\cal R$. We denote the automorphisms
by $\alpha_t: {\cal R} \rightarrow {\cal R}$, with $t$ real.   Let
us fix a concrete von Neumann algebra $\cal R$ on a Hilbert space
$H$, and a cyclic and separating vector $|\Psi\rangle$ in $H$.
Consider the operator $S$
defined by
\f
	S A |\Psi\rangle = A^\star |\Psi\rangle.    \label{S}
\ff
One can show that $S$ admits a polar decomposition
\f
	S = J \Delta^{1/2}					\label{s}
\ff
where $J$ is antiunitary, and $\Delta$ is a self-adjoint, positive
operator.  The Tomita-Takesaki theorem \cite{tt} states that the
map $\alpha_t: {\cal R} \rightarrow {\cal R}$ defined by
\f
	\alpha_t A =  \Delta^{-it} \ A\   \Delta^{it}
	\label{tomita}
\ff
defines a 1-parameter group of automorphisms of the algebra
$\cal R$.  This group is denoted the group of modular
automorphisms, or the modular group, of the state $\omega$ on
the algebra $\cal R$ and
will play a central role in the following.\footnote{The modular
group is usually defined with the opposite sign of $t$. We have
reversed the sign convention in order to make contact with
standard physics usage.}

Notice that the  Tomita-Takesaki theorem applies also to an
arbitrary faithful state $\omega$ over an abstract
$C^\star$-algebra $\cal
A$, since $\omega$ defines a representation of $\cal A$ via
the GNS construction, and thus a von Neumann algebra with a
preferred state.

An automorphism $\alpha_{inner}$ of the algebra $\cal R$ is
called an inner automorphism if there is a unitary element $U$ in
$\cal R$ such that
\f
		\alpha_{inner} A = U^\star A U.
\ff
Not all automorphisms are inner.   We may consider the following
equivalence relation in the family of all automorphism of $\cal
R$.  Two automorphisms $\alpha'$ and $\alpha'{}'$ are
equivalent when they are related by an inner automorphism
$\alpha_{inner}$, namely $\alpha'{}'=\alpha_{inner} \alpha'$, or
\f
	\alpha' (A)\ U  =   U\ \alpha'{}'(A),
\ff
for every $A$ and some $U$ in $\cal R$.  We denote the
resulting equivalence classes of automorphisms as outer
automorphisms, and the space of the outer automorphisms of
$\cal R$ as Out($\cal R$).   In general, the
modular group $\alpha_t$ (\ref{tomita}) is not a group of inner
automorphisms. It follows that in general $\alpha_t$ projects
down to a non-trivial
1-parameter group in Out($\cal R$), which we shall denote as
$\tilde\alpha_t$.  The Cocycle Radon-Nikodym \cite{al}
theorem states that two
modular automorphisms defined by two states of a von Neumann
algebra are inner-equivalent.   It follows that all states of a von
Neumann algebra determine the {\it same\,} 1-parameter group in
Out($\cal R$), namely $\tilde\alpha_t$ does not depend on
$\omega$.   In other words, a von Neumann algebra possesses a
{\it canonical\,} 1-parameter group of outer automorphisms.
This group plays a central role in the classification of the von
Neumann algebras;  in sec 5 we shall suggest a physical
interpretation for this group.

\vskip.3cm
{\it 2.2 The problem of the choice of the physical time in general
covariant theories}
\vskip.3cm

Let us return to physics. There are several open difficulties
connected with the treatment of the notion of time in general
covariant quantum theories, and it is important to distinguish
carefully between them.

General covariant theories can be formulated in the lagrangian
language in terms of evolution in a non-physical, fictitious
coordinate time.  The coordinate time (as well as the spatial
coordinates) can in principle be discarded from the formulation of
the theory without loss of physical content, because results of
real gravitational experiments are always expressed in
coordinate-free form.  Let us generically denote the fields of the
theory as $f_A(\vec x, x^o),\ A=1...N$.  These include for instance
metric field, matter fields, electromagnetic field, and so on, and
are subject to equations of motion invariant under coordinate
transformations. Given a solution of the equations of motion
\f
	f_A = f_A(\vec x, x^o), 				\label{fapar}
\ff
we cannot compare directly the quantities $f_A(\vec x, x^o)$ with
experimental data.  Results of experiments, in fact, are expressed
in terms of physical distances and physical time intervals, which
are functions of the various fields (including of course the metric
field) independent from the coordinates $\vec x, x^o$.  We have to
compute coordinate independent quantities out of the quantities
$f_A(\vec x, x^o)$, and compare these with the experimental
data.\footnote{Cases of clamorous oversight of this
interpretation rule are known, as an unfortunate determination of
the Earth-Moon distance, in which a meaningless {\it
coordinate\,} distance survived in the literature for a while.}

The strategy employed in experimental gravitation, is to
use concrete physical objects as clocks and as spatial references.
Clocks and other reference system objects are concrete physical
objects also in non generally covariant theories; what is new in
general covariant theories is that these objects cannot be taken
as independent from the dynamics of the system, as in non
general-covariant physics.  They must be components of the
system itself.\footnote{There is intrinsically no way of
constructing a physical clock that is not affected by the
gravitational field.}  Let these "reference system objects" be
described by the variables $f_1... f_4$ in the theory.  We are more
concerned here with temporal determination than with space
determination.  Examples of physical clocks are: a laboratory
clock (the rate of which depends by the local gravitational field),
the pulsar's pulses, or an arbitrary combination of solar system
variables, these variables are employed as independent variables
with respect to which the physical evolution of any other variable
is described  \cite{tradition,observable}.  In the theoretical
analysis of an experiment, one typically first works in terms of
an (arbitrary) coordinate system $\vec x, x^o$, and then one
compares a solution of the equations of motion, as (\ref{fapar})
with the data in the following way.  First we have to locally solve
the coordinates $\vec x, x^o$ with respect to quantities $f_1...
f_4$ that represent the physical objects used as clocks and as
spatial reference system
\f
	f_1(\vec x, x^o) ... f_4(\vec x, x^o)\  \rightarrow \
	\vec x(f_1, ..., f_4), x^o(f_1, ..., f_4) 	\label{inversion}
\ff
and then  express the rest of the remaining fields ($f_i\ \
i=5...N$) as  functions of $f_1 ... f_4$
\f
	f_i(f_1, ..., f_4) = f_i(\vec x(f_1, ..., f_4), x^o(f_1, ..., f_4)).
								\label{inversion2}
\ff
If, for instance, $F(\vec x, t)$ is a scalar, then for every
quadruplet of numbers $f_1 ... f_4$, the quantity $F_{f_1...f_4}
=F(f_1, ..., f_4)$ can be compared with experimental data.  This
procedure is routinely performed in any analysis of experimental
gravitational data -- the physical time $f_4$ representing
quantities as the reading of the laboratory clock, the counting of
a pulsar's pulses, or an arbitrary combination of solar system
observed astronomical variables.

The role of the coordinates (and in particular of the time
coordinate) can be clarified by means of a well known analogy.
The coordinates have the
same physical status as the arbitrary parameter $\tau$ that we
use in order to give a manifestly Lorentz covariant description of
the motion of a relativistic particle.  The physical motion of a
particle (non-relativistic as well as relativistic) is described by
the three functions of one variable
\f
	\vec X = \vec X(t),				\label{unpar}
\ff
where $\vec X$ is the position in a coordinate system and $t$
the corresponding time.   We can introduce an arbitrary parameter
$\tau$ along the trajectory, and describe the motion (\ref{unpar})
in the parametrized form
\begin{eqnarray}
	\vec X & = & \vec X(\tau)	  		\nonumber \\
	t & = & t(\tau);				\label{par}
\end{eqnarray}
the advantage of this description is that the dynamics can be
formulated in manifestly Lorentz covariant form.  In fact, we can
put $X^\mu(\tau)= (\vec X(\tau), t(\tau))$, and show that $X^\mu$
satisfies the manifestly Lorentz covariant dynamics generated by
the action $S[X]=\int d\tau \sqrt{\dot X^\mu \dot X_\mu}$.   If we
are given a solution of the equations of motion, we cannot
compare directly the numbers $X^\mu(\tau)= (\vec X(\tau),
t(\tau))$ with experimental observations.  We should recover the
physical motion (\ref{unpar}) by "deparametrising" (\ref{par}),
that is by picking the time variable
\f
	t = X^o						\label{choice}
\ff
out of the four variables $x^\mu$, solving $\tau$ with respect to
the time variable $t$
\f
	t(\tau) = X^o(\tau) \rightarrow \tau(t),
\ff
and replacing $\tau$ with $t$ in the rest of the equations
(\ref{par})
\f
	\vec X(t) = \vec X(\tau(t)).
\ff
This way of getting rid of the parameter $\tau$ is exactly the
same as the way we get rid of the four coordinates in
(\ref{inversion}--\ref{inversion2}).   Furthermore, the equations
of motions generated by the action $S[X]$ are invariant under
reparametrisation of $\tau$, as the generally covariant equations
are invariant under reparametrisation of the four coordinates.

Notice that $t$, as defined in (\ref{choice}),  is not the only
possible time variable.  Any linear combination $t' =
\Lambda^o_\mu X^\mu$, where
$\Lambda$ is a matrix in the Lorentz group, is another possible
time variable.  We know, of course,  the physical meaning of
this abundance of time variables: they represent the different
physical times of different Lorentz observers in relative motion
one with respect to the other.  Given a choice of one of the
Lorentz times, we can construct a conventional
``non-parametrised" dynamical system with a hamiltonian that
describes the evolution of the particle in that
particular time variable.

In the case of a general covariant theory, we also select one of
the lagrangian variables as time and express the evolution of
the system in such a time, say
\f
	t = f_4
\ff
in the example above.  And we also have a large freedom in this
choice. The variable selected is sometimes denoted internal time
or physical time.

Let us now illustrate how this strategy is implemented in the
canonical formalisms. In the hamiltonian formalism, general
covariance implies that the hamiltonian vanishes weakly.  The
most common way of constructing the canonical formulation of
general relativity is by starting from an ADM spacelike surface,
and identifying canonical variables as values of fields on this
surface.  While   powerful, this approach is very
unfortunate from the conceptual side, since it is based on the
fully non-physical notion of a ``surface" in spacetime.  This
procedure breaks {\it explicit\/} four-dimensional covariance,
and thus it has lead to the popular wrong belief that the canonical
formalism is intrinsically non-generally covariant.  An
alternative approach exists in the literature, and is based on the
covariant interpretation of the phase space as the space of the
solutions of the equations of motion.   This approach is known in
mechanics since the nineteen century, it shows that the canonical
theory is as "covariant" as the covariant theory, and avoids the
interpretational problems raised by the ADM approach (see for
instance ref. \cite{luca} and the various references there). Let
$\Gamma_{ex}$ be the space of solutions of the generally
covariant equations of motion.  On $\Gamma_{ex}$,
a degenerate symplectic structure is defined by the equations
of motion themselves.    The degenerate directions of this
symplectic structure integrate in orbits. One can show that these
orbits are the orbits of the 4-dimensional diffeomorphisms,
and thus all solutions of the equation that are in the same orbit
must be identified physically.  The space of these
orbits, $\Gamma$, is a symplectic space: the physical phase
space of the theory.   One can show that $\Gamma$ is isomorphic
to the reduced ADM phase space, proving in this way that the
reduced ADM phase space is a fully 4-dimensionally covariant
object independent from the unphysical ADM hypersurface
generally used for its construction.

The quantities that can be compared with
experimental data are the smooth functions on this space: these
are, by construction, all the coordinate independent quantities
that can be computed out of the solutions of the Einstein
equations.

For instance, for every quadruplet of numbers $f_1 ... f_4$ the
quantity $ F_{f_1 ... f_4}$ considered above is a function on
$\Gamma$, which expresses the value of the observable $F_{f_1
... f_4}$ on the various solutions of the equations of motion.
The quantity $F_{f_1 ... f_4}$ is well defined on $\Gamma$
because it is constant along the orbits in $\Gamma_{ex}$, due to
the fact that the coordinates $(\vec x, x^o)$ have been solved
away.  From now on, we focus on the number that
determines the time instant, say $t = f_4$.  By (fixing $f_1...f_3$
and) changing $t$ in
$F_{f_1 ... f_3, t}$ , we obtain a one parameter family of
observables $F_{t}$,
expressing the evolution of this observable in the clock-time
defined by $f_4$.   Note that we have defined  $F_{t}$ as a
function on $\Gamma$ only implicitly: in general, writing this
function explicitly is a difficult problem, which may amount to
finding the general solution of the equations of motion.   In the
canonical formulation of a generally covariant theory, therefore,
there is no hamiltonian defined on the phase space, but the phase
space is a highly non-trivial object, and the smooth functions on
the phase space, namely the generally covariant observables,
contain dynamical information implicitly.

Let us now consider the formal structure of a generally covariant
{\it quantum\,} theory.  The set ${A} = C^\infty(\Gamma)$ of the
real
smooth functions on $\Gamma$, namely of the physical
observables, form an abelian multiplicative algebra.   If we
regard the dynamical system as the classical limit of a quantum
system, then we must interpret the observables in
$\cal A$ as classical limits of non-commuting quantum
observables.   We assume that the
ensemble of the quantum observables form a non-abelian
$C^\star$-algebra, which we denote as $\cal A$.   Since
$\Gamma$ is a symplectic space, ${A}$ is a non-abelian algebra
under the
Poisson bracket operation.  We can view this Poisson structure as
the classical residue of the non-commuting quantum structure,
and thus assume that the quantum algebra $\cal A$ is a
deformation of  (a subalgebra of) the classical Poisson algebra.
Since we are dealing with a generally covariant theory, no
hamiltonian evolution or representation of the Poincare' group are
defined in $\cal A$.   A time evolution is only determined by the
dependence of the observables on clock times. For instance, in the
example above $F_{t}$ has an explicit dependence on $t$.

In using the strategy described above for representing the
physical time evolution in a generally covariant theory, however,
we encounter two difficulties.   The first difficulty is that in
general relativity no selection of an internal time is known,
which leads to a well defined fully non-parametrised dynamical
system.  Typically, in an arbitrary solution of the equations of
motion, an arbitrary internal time, as $t = f_4$, does not grow
monotonically everywhere, or equivalently, the inversion
(\ref{inversion}) can
be performed only locally.    In the classical context, we can
simply choose a new clock when the first goes bad, but the
difficulty becomes serious in the quantum context, for two
reasons.  First, because we would like to define the quantum
operator on (at least a dense subset of) the entire Hilbert space,
and we may have troubles in dealing with quantities that make
sense on some states only. This is the mathematical counterpart
of the physical fact that in a generic superposition of geometries
a physical clock may stop and run backward on some of the
geometries.  Second, because the corresponding quantum
evolution lacks unitarity, and the problem is open whether this
fact jeopardizes the consistency of the interpretation of the
quantum formalism.   Indeed, this lack of unitarity of the
evolution
generated by an arbitrary internal time has frequently been
indicated as a major interpretational problem of
quantum gravity.  On the other hand, it has also been argued (for
instance in ref. \cite{time}) that a consistent interpretation of
the quantum formalism is still viable also in the general case in
which a preferred unitary time evolution is lacking.  In the
present work, we do {\it not\,} concern ourselves with this
problem.   We simply assume that a generally covariant quantum
theory can be consistently defined.

The second difficulty, and the one we address in this paper, is
that of an embarrassment of riches.  The problem is the
following.  If the formalism of the theory remains consistent
under an essentially arbitrary choice of the evolution's
independent variable, then what do we mean by ``the" time?
Namely, what is it that characterizes a time variable as such?
How do we relate the freedom of choosing any variable as the
time variable, which a general covariant theory grants us, with
our (essentially non-relativistic, unfortunately) intuition of a
preferred, and, at least locally, unique parameter measuring the
local time flow?  Thus, the problem that we consider in this
paper is:  What is it that singles out a particular flow as the
physical time?

The elusive character of the notion of time and
the difficulty of capturing its stratified meanings for purely
mechanical theories are well known, and have been pointed out
and extensively discussed in the literature, even in the more
limited context of non-relativistic theories \cite{libri}.  The
discovery that the world is described by a general covariant
theory makes this problem much more severe.

\vskip.9cm
\vfill\pagebreak

{\bf 3. The time flow is determined by the thermal state}
\vskip.3cm
{\it 3.1 The modular group as time}
\vskip.3cm

The hypothesis that we explore in this paper is that the notion of
a preferred ``flowing" time has no mechanical meaning at
the quantum generally covariant level, but rather has
thermodynamical origin.   The idea that thermodynamics and the
notion of a time flow are deeply intertwined is as old as
thermodynamics itself, and we shall not elaborate on it here.  To
be clear, what we intend to ascribe to thermodynamics is not the
{\it versus\,} of the time flow.   Rather, it is the time flow
itself, namely the specification of which one is the independent
variable that plays the physical role of time, in a fundamental
general covariant theory.

By thermodynamical notion we mean here a notion that makes
sense on an ensemble, or, equivalently, on a {\it single\,} system
with many degrees of freedom, when we do not have access to its
full microscopic state, but only to a number of macroscopic
coarse-grained variables, and therefore we are forced to describe
it in terms of the distribution $\rho$ of the microscopic states
compatible with the macroscopic observations, in the sense of
Gibbs \cite{gibbs}.   Notice that in
field theory we are always in such a thermodynamical context,
because we cannot perform infinite measurements with infinite
precision.  The observation that a fundamental description of the
state of a field system is always incomplete, and therefore
intrinsically thermodynamical, in the sense above, is an
important ingredient of the following discussion.

In the context of a conventional non-generally covariant quantum
field theory, thermal states are described by the KMS condition.
Let us recall this formalism.  Let $\cal A$ be an
algebra of quantum operators $A$; consider the 1-parameter
family of automorphisms of $\cal A$ defined by the time
evolution
\f
	\gamma_t A = e^{i t H/\hbar}\ A\ e^{-i t H/\hbar}  \label{tf}
\ff
where $H$ is the hamiltonian.  From now on we put $\hbar=1$. We
say that a state $\omega$ over $\cal A$ is a
Kubo-Martin-Schwinger (KMS) state (or satisfies the KMS
condition) at inverse temperature $\beta = 1/k_{b}T$ ($k_b$ is
the Boltzmann constant and $T$ the absolute temperature), with
respect to $\gamma_t$, if the
function
\f
	f(t) = \omega(B (\gamma_{t} A))
\ff
is analytic in the strip
\f
	0 < {\rm Im\ } t < \beta
\ff
and
\f
	\omega((\gamma_t A) B) = \omega(B (\gamma_{t+i\beta}
A)).
\ff
Haag, Hugenholtz and Winnink \cite{hhw} have shown that this
KMS condition
reduces to the well known Gibbs condition
\f
	\omega= N e^{-\beta H}				\label{gibbs}
\ff
in the case of systems with finite number of degrees of freedom,
but can also be extended to systems with infinite degrees of
freedom. They have then postulated that the KMS condition
represents the correct physical extension of the Gibbs postulate
(\ref{gibbs}) to infinite dimensional quantum systems.

The connection between this formalism and the Tomita-Takesaki
theorem is surprising.  Any faithful state is a KMS
state (with inverse temperature $\beta =1$) with respect to the
modular automorphism $\alpha_t$ it itself generates [14].  Therefore,
in a non-generally covariant theory, an equilibrium state is a
state whose modular automorphism group is the time translation
group (where the time is measured in units of
$\hbar/k_{b}T$).  Thus, as we noticed in the introduction
in the
context of classical mechanics, an equilibrium quantum thermal
state contains all the information on the dynamics which is
contained in the hamiltonian (except for the constant
factor $\beta$, which depends on the unit of time employed).   If
we take into account the fact that the universe around us {\it
is\,} in a state of non-zero temperature, this means that the
information about the dynamics can be fully replaced by the
information about the thermal state.

To put it dramatically, we could say that if we knew
the thermal state of the fields around us (and we knew the
full kinematics, namely the specification of all the
dynamical fields), we could then throw away the full
Standard Model {\it lagrangian\ } without loss of information.

Let us now return to generally covariant quantum theories.  The
theory is now given by an algebra $\cal A$ of generally covariant
physical operators, a set of states $\omega$, over $\cal A$, and
no additional dynamical information.  When we consider a
concrete physical system, as the physical fields that surround
us, we can make a (relatively small) number of physical
observation, and therefore determine a (generically impure) state
$\omega$ in which the system is.    Our problem is to
understand the origin of the physical time flow, and our working
hypothesis is that this origin is thermodynamical.  The set of
considerations above, and in particular the observation that in a
generally covariant theory notions of time tend to be state
dependent, lead us to make the following hypothesis.
\begin{quote}
\em The physical time depends on the state.\  When the system
is in a state $\omega$, the physical time is given by the modular
group $\alpha_t$ of $\omega$.
\end{quote}
The modular group of a state was defined in eq.(\ref{tomita})
above. We call the time flow defined on the algebra of the
observables by the modular group as the {\it thermal time}, and
we denote the hypothesis above as the thermal time hypothesis.

The fact that the time is determined by the state, and therefore
the system is always in an equilibrium state with respect to the
thermal time flow, does not imply that evolution is frozen, and
we cannot detect any dynamical change.  In a quantum system
with an infinite number of degrees of freedom, what we generally
measure is the effect of small perturbations around a thermal
state.   In conventional quantum field theory we can extract
all the information in terms of vacuum expectation values of
products of fields operators, namely by means of a single
quantum state $|0\rangle$. This was emphasized by Wightman. For
instance, if $\phi$ is a scalar field, the propagator (in a fixed
space point $\vec x$) is given by
\f
	F(t) 	= \langle0|\phi(\vec x,t)\phi(\vec x,0)|0\rangle
		=  \omega_o(\gamma_t(\phi(\vec x,0))\,\phi(\vec
x,0)),
\ff
where $\omega_o$ is the vacuum state over the field algebra, and
$\gamma_t$ is the time flow (\ref{tf}).  Consider a generally
covariant quantum field theory.  Given the quantum algebra of
observables $\cal A$, and a quantum state $\omega$, the modular
group of $\omega$ gives us a time flow $\alpha_t$.  Then, the
theory describes physical evolution in the thermal time in terms
of amplitudes of the form
\f
	F_{A, B}(t) = \omega(\alpha_t(B)\,A)
\ff
where $A$ and $B$ are in $\cal A$. Physically, this quantity is
related to the amplitude for detecting a quantum excitation of
$B$ if we prepare $A$ and we wait a time $t$ -- ``time" being
the thermal time determined by the state of the system.

In a general covariant situation, the thermal time is the only
definition of time available. However, in a theory in which a
geometrical definition of time independent from the thermal time
can be given, for instance in a theory defined on a Minkowski
manifold, we have the problem of relating geometrical time and
thermal time.  As we shall see in the examples of the following
section, the Gibbs states are the states for which the two time
flow are proportional.  The constant of proportionality is the
temperature.  Thus, within the present scheme
the temperature is interpreted as the ratio between thermal
time and geometrical time, defined only when the second is
meaningful.\footnote{It has been suggested, for instance by
Eddington \cite{edi}, that any clock is necessarily a
thermodynamical clock.  It is tempting to speculate that in a fully
quantum generally covariant context, namely in a Planck regime,
the only kind of clock that may survive are quantum thermal
clock, as for instance decay times, which naturally measure the
thermal time.}

We believe that the support to the thermal time hypothesis comes
from analyzing its consequences and the way this hypothesis
brings disconnected parts of physics together.  In the following
section, we explore some of these consequences. We will
summarize the arguments in support the thermal time hypothesis
in the conclusion.

\vskip.9cm
\vfill\pagebreak

{\bf 4. Consequences of the hypothesis}
\vskip.3cm

{\it 4.1 Non relativistic limit}
\vskip.3cm

If we apply the thermal time hypothesis to a non-generally
covariant system in thermal equilibrium, we recover immediately
known physics. For simplicity, let us consider a system with a
large, but finite number of degrees of freedom, as a quantum gas
in a box.

The quantum state is then given by the Gibbs density matrix
\f
	\omega = N e^{-\beta H}			\label{gib}
\ff
where $H$ is the hamiltonian, defined on an Hilbert space $\cal
H$, and
\f
	N^{-1} = \tr\, [ e^{-\beta H} ].
\ff
The modular flow of $\omega$ is
\f
 	\alpha_t  A  = e^{i\beta t H}\ A\ e^{-i\beta t H},    \label{flow}
\ff
namely it is the time flow generated by the hamiltonian,
with the time rescaled as $t\rightarrow \beta t$.
This can be proven directly by checking the KMS condition between
$\omega$ and $\alpha_t$, and invoking the uniqueness of the
KMS flow of a given state.

Alternatively, one can also explicitely construct
the modular flow from $\omega$, in terms of the
operator $S$ as in eqs. (\ref{S},\ref{s},\ref{tomita}).
 Let us briefly describe such
construction, since it has the merit of displaying the physical
meaning of the mathematical quantities appearing in the
formulation of the Tomita-Takesaki theorem.

The Tomita-Takesaki construction cannot be applied directly, because
the state $\omega$ is not given by a vector of the Hilbert space.
We have first to construct a new representation of the
observables algebra in which $\omega$ is given by a vector
$|0\rangle$, using the GNS construction.  A shortcut for
constructing this representation was suggested in ref.\cite{hhw}:
The set of the Hilbert-Shmidt density matrices on $\cal H$,
namely the operators $k$ such that
\f
	\tr\, [k^\star k] \ <\ \infty
\ff
forms a Hilbert space, which we denote as $\cal K$. We denote an
operator $k$ in $\cal H$ as $|k\rangle$ when thought as a vector
in $\cal K$.   We can construct a new representation of the
quantum theory on (a subspace of) this Hilbert space $\cal K$.
This will be a (reducible) representation in which the thermal
Gibbs state is given by a pure vector.   The operator
\f
	k_o = \omega^{1/2}
\ff
is in $\cal K$, and we denote it as $|k_o\rangle$.  The operator
algebra of the quantum theory acts on $\cal K$ as follows. If $A$
is a quantum operator defined in $\cal H$, then we can write
\f
	A |k \rangle = |A k\rangle.
\ff
Which is again in $\cal K$.  By taking $|k_o\rangle$ as the cyclic
state, and acting on it with all the $A$'s in the observables
algebra we generate the new representation that we are
searching.   The state $|k_o\rangle$ plays the role of a ``vacuum
state" in this representation, but we can either increase or
decrease its energy, as must be for a thermal state.

Note the peculiar way in which the time translations group
$U_{\cal K}(t)$ act on this representation.  The state
$|k_o\rangle$ is of course time invariant
\f
	U_{\cal K}(t) |k_o\rangle = |k_o\rangle.
\ff
A generic state is of the form $|A k_o\rangle$; its time
translated is
\f
	U_{\cal K}(t) | A k_o\rangle = | \gamma_t(A) k_o\rangle
	= | e^{itH}A e^{-itH} k_o\rangle
	= | e^{itH}A k_o e^{-itH} \rangle .
\ff
where we have used the time flow $\gamma_t$ given by the
(Heisenberg) equations of motion (\ref{tf}).
Therefore we have in general
\f
	U_{\cal K}(t) |k\rangle = |e^{itH} k e^{-itH} \rangle.
\ff

The interest of the representation defined above is due to the
fact that it still exists in the thermodynamical limit in which the
number of degrees of freedom goes to infinity. In this limit, the
expression (\ref{gib}) looses sense, since the total energy of an
infinity extended thermal bath is infinite. However, the $\cal K$
representation still exists, and includes all the physical states
that are formed by finite excitations around (``over" and ``below"!)
the thermal state $|k_o\rangle$.   Here, we are interested in this
representation as a tool for constructing the modular group, to
which we now return.

The modular group depends on the operator $S$, defined in
eq.(\ref{s}). Here we have
\f
	S A |k_o\rangle = A^\star  |k_o\rangle = |A^\star k_o\rangle
{}.
\ff
or
\f
	S A  e^{-\beta H/2}  = A^\star e^{-\beta H/2},
\ff
because
\f
	k_o = \omega^{1/2} =  N^{1/2} e^{-\beta H/2}.
\ff
It is then easy to check that the polar decomposition  $S = J
\Delta^{1/2} $ is given by
\f
	J  | k \rangle= | k^\star \rangle
\ff
and
\f
	\Delta^{1/2}|k\rangle
	= |e^{-\beta/2  H} k e^{\beta/2  H} \rangle.
\ff
In fact we have
\begin{eqnarray}
	S A |k_o\rangle
	&=& J \Delta^{1/2} |A  N^{1/2} e^{-\beta H/2}\rangle
\nonumber \\
	&=& J |e^{-\beta/2  H} A  N^{1/2} e^{-\beta
		H/2}e^{\beta/2  H}\rangle
\nonumber  \\
	&=& J |e^{-\beta/2  H} A  N^{1/2}\rangle
\nonumber  \\
	&=&  |  N^{1/2} A^\star e^{-\beta/2  H}\rangle
\nonumber  \\
	&=& A^\star |k_o\rangle.
\end{eqnarray}
The operator $J$ exchanges, in a sense, creation operators with
annihilation operators around the thermal vacuum, and therefore
contains the information about the splitting of the representation
between the states with higher and lower energy than the thermal
vacuum. The object that we are searching is the modular
automorphism group $\alpha_t$, which is given defined in
eq.(\ref{tomita}), as
\begin{eqnarray}
	\alpha_t A |k_o\rangle
	&=&  \Delta^{-it}\ A\  \Delta^{it} |k_o\rangle
\nonumber  \\
	&=&  \Delta^{-it}\ A \ |k_o\rangle
\nonumber  \\
	&=&    |e^{i\beta t H} A k_o e^{-i\beta t H}\rangle
\nonumber  \\
	&=&   |e^{i\beta t H} A e^{-i\beta t H} k_o \rangle.
\end{eqnarray}
Namely
\f
	\alpha_t  A  = e^{i\beta t H}\ A\ e^{-i\beta t H} .
\ff
So that we may conclude that the modular group and the time
evolution group are related by
\f
	\alpha_t    = \gamma_{\beta t}.
\ff
We have shown that the modular group of the Gibbs state is the
time evolution flow, up to a constant rescaling $\beta$ of the
unit
of time.   Thus, if we apply the thermal time postulate to the
Gibbs state (\ref{gib}), we obtain a definition of physical time
which is proportional to the standard non-relativistic time.

We also note the following suggestive fact \cite{statistical2}. In
a special relativistic system, a thermal state breaks Lorentz
invariance.  For instance, the average momentum of a gas at
finite temperature defines a preferred Lorentz frame. In other
words, a thermal bath is at rest in one Lorentz frame only.  If we
apply the postulate to such a state, we single out a preferred
time, and it is easy to see that this time is the Lorentz time of
the Lorentz frame in which the thermal bath is at rest.

\vskip.3cm
{\it 4.2 Classical limit}
\vskip.3cm

Let us return to a fully general covariant system, and consider a
state $\omega$, and an observable $A$.  From the definition of
the modular group we have
\f
	\alpha_t A  = e^{-it ln\Delta}\ A\  e^{it ln\Delta} ,
\ff
and therefore
\f
	 \dot A \equiv
	\left. {d\over dt} \alpha_t A\right|_{t=0}
	= i [A, ln\Delta].					\label{com}
\ff
Note, from the previous subsection, that
\f
	[A,\Delta] = [A,\omega].
\ff
Let us now consider the classical limit of the theory. In this
limit, if we replace observables, as well as density matrices,
with functions on the phase space, then commutators are replaced
by Poisson brackets.  Let us denote the classical observable
corresponding to the operator $A$ as $A$, and the classical
density matrix that approximates the quantum density matrix
$\omega$, by $f$ and $\omega$.  The formal classical limit can be
obtained by the standard replacement of commutators with
Poisson brackets. We have then that in the classical limit
the thermal time flow is defined by the equation
\f
	 {d\over dt} f  = \{-\ln\rho\, f\},
\ff
namely
\f
	 {d\over dt} f  =  \{H, f\},			\label{hamilton}
\ff
where $H$ is defined as $H= -\ln\rho$, or
\f
	 \rho = e^{-H}.					\label{gibbs2}
\ff
Thus we obtain the result that there is a classical hamiltonian
$H$ that generates the Hamilton evolution, and that the state
$\rho$ is related to this hamiltonian by the Gibbs relation.  The
Hamilton equations (\ref{hamilton}) and the Gibbs postulate
(\ref{gibbs2}), are both contained in the Tomita-Takesaki relation
(\ref{tomita}).   In other words, Hamilton equations and
Gibbs postulate can be derived from the thermal time hypothesis.

These relations hold in general for a generally
covariant theory;  however, we recall that the Hamiltonian $H$
plays a quite different role than in a non-relativistic theory: it
does not determine the Gibbs state, but, rather, it is
determined by any thermal state.

\vskip.3cm
{\it 4.3 Rindler wedge, Unruh temperature and Hawking radiation}
\vskip.3cm

Consider a free quantum field theory on Minkowski space, in its
(zero temperature) vacuum state $|0\rangle$.  Consider an
observer $O$ that moves along a rectilinear and uniformly
accelerated trajectory with acceleration $a$; say along the
trajectory
\begin{eqnarray}
	x^o(s) & = & a^{-1} \sinh(s),     \nonumber \\
	x^1(s) & = & a^{-1} \cosh(s),     \nonumber \\
	x^2  \  \ = & x^3  &  = \ \  0.
\end{eqnarray}
Because of the causal structure of Minkowski space, $O$ has only
access to a subspace of Minkowski space, the Rindler wedge
$R$, defined by
\f
	x^1 > |x^o|.
\ff
Accordingly, he can only describe the system in terms of the
algebra of observables ${\cal A}_R$ which is the subalgebra of
the full fields algebra $\cal A$ of the quantum field theory,
obtained by restricting the support of the fields to the Rindler
wedge.

The Rindler wedge is left invariant by the Lorentz boosts in the
$x^1$ direction.  Let $k_1$ be the generator of these boosts in
the Lorentz group, and $K_1$ be the generator of these boosts in
the representation of the Poincare' group defined on the Hilbert
space of the theory.  Clearly the 1-parameter group generated by
$K_1$ leaves ${\cal A}_R$ invariant.

The Lorentz transformation
\f
	\Lambda(\tau) = \exp\{\tau \, a \, k_1\}
\ff
carries $O$ along its trajectory, $\tau$ being the proper time of
$O$.  This fact suggests that the unitary group of transformations
\f
	\gamma_\tau A
	= e^{i\,\tau\,a\,K_1}\ A\ e^{-i\,\tau\,a \,K_1}
\ff
can be interpreted as the physical time characteristic of the
observer $O$.  Let us approach the problem of the determination
of the physical time from the point of view of the thermal time
postulate.

The restriction of the state $|0\rangle$ to the algebra ${\cal
A}_R$ is of course a state on ${\cal A}_R$, and therefore it
generates a modular group of automorphisms $\alpha_t$ over
${\cal A}_R$.   Bisognano and Wichmann \cite{bisognano} have
proven that the modular group of $|0\rangle$ over ${\cal A}_R$ is
precisely
\f
	\alpha_t = \beta_{2 \pi\, a^{-1}\, t}.
\ff
Therefore, the thermal time of the system that the observer can
reach is proportional to the time flow determined geometrically
by its proper time.

In this case we have two independent and compatible
definitions of time flow in this system: the thermal time flow
$\alpha_t$ and the flow $\beta_\tau$ determined by the proper
time.   We have obtained
\f
	t =  { 2 \pi\over a} \tau.
\ff
We now interpret temperature as the ratio between the thermal
time and the geometrical time, namely $t=\beta\tau$. We obtain
$\beta = { 2 \pi\over a}$,  namely
\f
	T = {1\over k_b\beta} =  {a\over 2\pi k_{b}},
\ff
which is the Unruh temperature  \cite{unruh}, or the temperature
detected by a thermometer moving in $|0\rangle$ with
acceleration $a$.

An important aspect of this derivation of the time flow via the
modular group $\alpha_t$ of the algebra ${\cal A}_R$ is given by
the fact that the result depends only on the trajectory of the
observer (which determines $A_R$), and does not require any a
priori choice of coordinates on the Rindler Wedge.

Following Unruh's initial suggestion \cite{unruh}, this
construction can be immediately generalised to the
Schwarzschild solution -- an observer at rest in the
Schwarzschild coordinates undergoes a constant acceleration
pointing away from the black hole -- and the result allows one to
derive the Hawking temperature \cite{hawking}.  We will not
elaborate on this here.

\vskip.3cm
{\it 4.3 State independent notion of time: the canonical subgroup
of $Out\, A$}
\vskip.3cm

Finally, let us point out an intriguing aspect of the definition of
time that we have consider.  As shown in section 2, while
different states over a von Neumann algebra $\cal R$ define
distinct
time flows, still it is possible to define a state--independent
flow, in the following sense.   In general, the modular flow is not
an inner automorphism of the algebra, namely, there is no
hamiltonian in $\cal R$ that generates it.  Due to the
Cocycle
Radon-Nikodym  theorem, however, the difference between
two modular flows is always an inner automorphism. Therefore,
whatever state one starts with, the modular flow projects down to
the same 1--parameter group of elements of the group of outer
automorphisms $Out\, A$.  This flow is canonical:
it depends only on the algebra itself.   Von Neumann algebras,
indeed, are classified by studying this canonical flow.

In conventional field theories, where we are independently
provided with a notion of time evolution, the only consequence of
this observation is that we must be sure to choose the correct
Type of von Neumann algebra to start with (Type III, as argued in
\cite{haag}, and elsewhere).    However, in a generally covariant
quantum context
this observation provides us with the possibility of capturing a
fully state independent (rather abstract) notion of time evolution.
In fact, it is the algebra itself that determines the allowed time
flows.   In this abstract sense, a von Neumann algebra is
intrinsically a ``dynamical" object.

Furthermore, as the Rindler example points out, the same state
may give rise to different time flows when restricted to
different subalgebras of the full observables algebra of the
theory.  Since in general a subalgebra of a von Neumann algebra is
not isomorphic to the full algebra, the general features of the
time evolution may change substantially by focussing on different
observables subalgebras, that is by probing the theory in
different regimes.\footnote{It is tempting to suggest that
different ``phases" of a generally covariant quantum field theory,
perhaps different ``phases" in the history of the universe could be
read as observations of different observable subalgebras on the
same state, and that the general features of the flows can differ
substantially from phase to phase, from a fundamental
timelessness of the Planck--phase, to the present
time flow. We do not know how this idea could be
concretely implemented in a sensible theory.}

\vskip.9cm
\vfill\pagebreak

{\bf 5. Conclusions}
\vskip.3cm

The hypothesis that we have put forward in this paper is that
physical time has a thermodynamical origin.  In a quantum
generally covariant context, the physical time is determined by
the thermal state of the system, as its modular flow
(\ref{tomita}).

The main indications in support this hypothesis are the following
\begin{itemize}
\item {\em Non-relativistic limit}. In the regime in which we
may disregard the effect of the relativistic gravitational field,
and thus the general covariance of the fundamental theory,
physics is well described by small excitations of a quantum field
theory around a thermal state
$|\omega\rangle$. Since  $|\omega\rangle$ is a KMS state of the
conventional hamiltonian time evolution, it follows that the
thermodynamical time defined by the modular flow of
$|\omega\rangle$ is precisely the
physical time of non relativistic physics.
\item {\em Statistical mechanics of gravity}. The statistical
mechanics of full general relativity is a surprisingly unexplored
area of theoretical physics \cite{statistical}.  In reference
\cite{statistical} it is shown that the classical limit of the
thermal time hypothesis allows one to define a general covariant
statistical theory, and thus a theoretical framework for the
statistical mechanics of the gravitational field.
\item {\em Classical limit; Gibbs states}. The Hamilton
equations, and the Gibbs postulate follow immediately from the
modular flow relation (\ref{tomita}).
\item  {\em Classical limit; Cosmology}.  We refer to
\cite{statistical2}, where it was shown that (the classical limit
of) the thermodynamical time hypothesis implies that the
thermal time defined by the cosmic background radiation is
precisely the conventional Friedman-Robertson-Walker time.
\item {\em Unruh and Hawking effects}. Certain puzzling aspects
of the relation between quantum field theory, accelerated
coordinates and thermodynamics, as the Unruh and Hawking
effects, find a natural justification  within the scheme presented
here.
\item  {\em Time--Thermodynamics relation}.   Finally, the
intimate intertwining between the notion of time and
thermodynamics has been explored from innumerable points of
view \cite{libri}, and need not be expanded upon in this context.
\end{itemize}

The difficulties of finding a consistent interpretation of a
general covariant quantum field theory are multi-fold.  In this
work, we have addressed one of these difficulties: the problem of
relating the ``timelessness" of the fundamental theory with the
``evidence" of the flow of time.  We have introduced a tentative
ingredient for the solution of this problem, in the form of a
general relation between the thermal state of a system and a
1-parameter group of automorphisms of the observables algebra,
to be interpreted as a time flow.

Since we have not provided a precise definition of ``physical
time", besides its identification with the non-relativistic time
and the Lorentz times in the non-(general) relativistic limit, we
have deliberately left a certain amount of vagueness in the
formulation of the thermal time hypothesis.  In fact, in a quantum
or thermal general covariant context, the problem is precisely to
understand what do we want to mean by time, or whether there is
a relevant structure that reduces to the non-relativistic time.
The thermal time hypothesis is thus the suggestion of taking the
modular flow as the relevant generalization of the
non-relativistic time.  This generalization allows us to embrace
in a unitary perspective a large variety of puzzling aspects of
general covariant physics.

The consequences of this hypothesis can be explored in a variety
of situations.  For instance, as the restriction of the observables
algebra to the Rindler wedge determines a time flow, similarly
the restriction to a different fixed region of spacetime
defines a corresponding time flow.  It would be interesting to
compute this flow, which, unlike the Rindler case, will not have
any obvious geometrical interpretation, and study its physical
interpretation.

We leave a large number of issues open.  It is not clear to us, for
instance, whether one should consider all the states of a general
covariant quantum system on the same ground, or whether some
kind of maximal entropy mechanism able to select among states
may make sense physically.

In spite of this incompleteness, we find that the number of
independent facts that are connected by the thermal time
hypothesis suggests that this hypothesis could be an ingredient of
the fundamental, still undiscovered, generally covariant quantum
theory.

\vskip1cm
We thank the Isaac Newton Institute, Cambridge, where this work
was begun, for hospitality.
This work was partially supported by the NSF grant PHY-9311465.

\end{document}